\def\year{2015}
\documentclass[letterpaper]{article}
\usepackage{modified}
\usepackage{times}
\usepackage{helvet}
\usepackage{courier}
\pdfoutput=1 
\usepackage{url}
\usepackage{latexsym}
\usepackage{amsmath}
\usepackage{multirow} 
\usepackage{algorithm}
\usepackage{balance}
\usepackage[pdftex]{graphicx} \pdfcompresslevel=9
\usepackage{algpseudocode}
\usepackage{subcaption}
\usepackage{url}

\begin{document}
\title{Automatic Construction of Evaluation Sets and Evaluation of Document Similarity Models in Large Scholarly Retrieval Systems}
\author{
Kriste Krstovski\textsuperscript{\small\textbf{\dag,\S}}, David A. Smith\textsuperscript{\small\textbf{\ddag}} and Michael J. Kurtz \textsuperscript{\small\textbf{\S}} \\
\textsuperscript{\small\textbf{\dag}}Harvard-Smithsonian Center for Astrophysics, Cambridge, MA \\
\textsuperscript{\small\textbf{\S}}College of Information and Computer Sciences, University of Massachusetts Amherst, Amherst, MA\\
\textsuperscript{\small\textbf{\ddag}}College of Computer and Information Science, Northeastern University, Boston, MA\\
kkrstovski@cfa.harvard.edu, dasmith@ccs.neu.edu, kurtz@cfa.harvard.edu\\
}
\maketitle

\begin{abstract}
Retrieval systems for scholarly literature offer the ability for the scientific community to search, explore and download scholarly articles across various scientific disciplines. Mostly used by the experts in the particular field, these systems contain user community logs including information on user specific downloaded articles. 
In this paper we present a novel approach for automatically evaluating document similarity models in large collections of scholarly publications. Unlike typical evaluation settings that use test collections consisting of query documents and human annotated relevance judgments, we use download logs to automatically generate pseudo-relevant set of similar document pairs. More specifically we show that consecutively downloaded document pairs, extracted from a scholarly information retrieval (IR) system, could be utilized as a test collection for evaluating document similarity models. Another novel aspect of our approach lies in the method that we employ for evaluating the performance of the model by comparing the distribution of consecutively downloaded document pairs and random document pairs in log space. Across two families of similarity models, that represent documents in the term vector and topic spaces, we show that our evaluation approach achieves very high correlation with traditional performance metrics such as Mean Average Precision (MAP), while being more efficient to compute. 
\end{abstract}

\section{Introduction}
Scholarly IR systems cover many scientific disciplines. While some of them focus on one or more closely related scientific fields ranging from the life sciences and biomedicine \cite{pubmed} to computer science \cite{acm} and electrical engineering \cite{ieee} to name a few, others cover a broader and not closely related range of disciplines, e.g. Google Scholar \cite{scholar.google.com}. Each of these, and many other scholarly literature retrieval systems, employs its own document similarity models and ranking algorithms. Mostly used by the experts in the particular field, these systems contain user community logs including information on downloaded articles\footnote{Throughout the paper we use the terms ``articles'' and ``documents'' interchangeably.}.

It is typically the case that the evaluation of document similarity models depends on having human annotated set of relevant documents which are costly to produce. In the past, the issue of creating such sets has been dealt with by combining the output of multiple retrieval systems and having human annotators manually go over the union of their results in order to create a set of relevant documents for the given document query. For instance, this approach has been traditionally employed as part of the Text REtrieval Conference (TREC) challenges \cite{Armstrong}. In the absence of human generated test collections, developers are often faced with the dilemma of how to decide which document similarity model to initially use with a new document collection where there is no evaluation set. In such instances creating a pseudo-relevant set of similar documents from existing human generated information is very beneficial. 

In this paper, we utilize information on downloaded documents from a scholarly IR system logs to automatically generate a pseudo-relevance set of similar documents. In the past query log information has been used to mine query specific pseudo-relevant document sets. For example, one of the earliest works on this topic explored clickthrough data \cite{Joachims}. In our approach we focus on utilizing the information on downloaded documents i.e. documents that the user has downloaded in a given time window. We use this information to automate the process of creating pseudo-relevant test collection for evaluating document similarity models in large collections of scholarly publications. As in the case with the clickthrough data, where the assumption is that the user is more likely to click on a document that is more relevant to the query than a random document, our approach is driven by the notion that the user is more likely to consecutively download a document more similar to the previously downloaded document than a random one. For a downloaded document A this notion allows us to treat documents that have been downloaded consecutively after document A as pseudo-relevant which in turn allows us to automatically generate test sets for evaluating document similarity models. The same notion of similarity has been explored by various collaborative filtering approaches for recommender systems. For example, information on consecutively purchased products for product recommender systems or information on consecutively loaned books for a library recommender, etc.

Document similarity models typically represent documents in a shared space where each document is represented independently using collection wide features that make similarity computation efficient. They range from the traditional vector space
 models, that represent documents using tf--idf values computed over a vocabulary \cite{Croft_book} to latent variable models, such as latent Dirichlet allocation (LDA) \cite{blei_LDA}, that represent documents as a mixture of topics. Finding similar documents involves computing similarity between document representations and ranking documents based on the similarity metric used.

In context of similarity, these metrics convey the perceptual similarity of two documents or more specifically the log of their perceptual similarity. Take for example the Jensen-Shannon (JS) divergence, the similarity metric used in the topic space. If two documents are completely dissimilar its divergence would be one which is equal to the log of the perceptual similarity whose value in such case is zero - the two documents are perceived as dissimilar. For two identical documents the JS divergence would be zero making the perceptual similarity equal to one. The same holds for other similarity metrics such as Cosine distance, which is typically used by the vector space models. When normalized by its relative range this metric follows the same interpretation. 

Within a ranked list of documents similar to a query document, models that perform better rank similar documents higher compared to dissimilar documents. This implies that the histogram plots across the set of values for the similar and dissimilar documents would observe distinct shapes. For the similar documents, the bulk of the histogram mass would be concentrated around the lower range of the scale. 
Expanding this notion to a set of consecutively downloaded document pairs and randomly generated document pairs we should also expect to see different shapes of the histogram plots across the two sets of similarity values. We hypothesize that this is due to the fact that the similarity values of the consecutively downloaded document pairs would be higher as these pairs were generated by an underlying human machinery i.e. users that have entered a particular query and have browsed the query results. Likewise, randomly generated documents would be less similar as they did not arise from human interaction. While there may be instances where such document pairs may indeed be similar, the bulk of their similarity values would be residing in the mode further out on the scale as defined by the similarity metric. 

We utilize this notion to come up with a new approach for evaluating document similarity models in large collections of scholarly publications. More specifically, we show that the slope of the curve obtained by dividing the histograms of the consecutively downloaded and randomly generated document pairs in log space has a very high correlation with the actual performance of the document similarity model as measured by a traditional IR metrics such as MAP. We refer to this approach as random histogram slope analysis (rHSA). Unlike a traditional IR evaluation metric, computing rHSA does not require traversing a ranked list for each query result, thus making it more efficient to compute.

\section{Similar vs. Dissimilar Document Pairs}

While many features could be used to compare documents such as scientific papers---citations, authors, institutional affiliations---we focus in this paper on topically similar documents.  In this context, topics are broadly defined as a set of words that have strong topical relatedness. Unlike randomly generated document pairs, topically similar document pairs have higher similarity as measured by the similarity metric used. The same notion, as we will see in the later chapters, holds for any document pairs that were generated through a human interaction - their topical relatedness would always be higher compared to randomly generated document pairs. Take for example consecutively downloaded document pairs - the set of document pairs that, as we will show in this paper, could be used as similar document pairs for constructing pseudo-relevance sets. The similarity between two consecutively downloaded document pairs due to the mere fact that were generated by users interaction with an IR system would always be higher than the similarity between two randomly generated document pairs.

Finding documents that are topically similar involves representing them in a shared space and comparing their representations using a particular metric. For example, the traditional term vector models use tf--idf values over a certain vocabulary to represent documents. In case of latent variable models, such as LDA, documents are represented as discrete probability distributions over a set of topics $T$. 

Regardless of the representation used, within the shared space, similar documents tend to be positioned close to each other versus dissimilar documents that are further apart. Across different model configurations the shared space could vary based on the number of dimensions. For example, in the shared topic space, the number of topics defines the dimensionality of the space. The same holds for the vector space where the number of tf--idf values used to represent documents defines the space dimensions. When retrieving similar documents different model configurations give different performance which is directly related to how the model positions documents in the shared space. 

One of the aspects of the model strength should be defined as how well does the model position documents that are similar compared to dissimilar documents. More powerful models should create a shared space such that for a given document $d_{i}$, represented as a topic vector $\theta_{i}$, its topically similar documents would be in its nearest proximity while the topically dissimilar documents would be positioned further away. 

Histograms are traditionally used as means to perform statistics and obtain visual information over the frequency of occurrence of values in a given range. They are also used as simple nonparametric approach for estimating the underlying probability density \cite{Bishop}. When computed over the similarity values of topically similar documents, histograms should observe distribution different from the distribution of topically dissimilar documents as their similarity values would be higher as they are positioned close to each other. 

\section{Automatic Evaluation of Document Similarity Models}
\label{sec:hsa}

Performance of document similarity models are typically evaluated using a test collection of query documents along with their sets of query relevant i.e. similar documents. Generating query relevant documents require human annotation and are traditionally being done by first generating a union of returned documents across multiple similarity retrieval systems. The set of returned results is then presented to a group of human annotators who manually judge their relatedness i.e. similarity to the query document. While lately there is an emerging trend to obtain human annotation through the use of crowdsourcing this still requires significant human effort on the developer side in creating the proper infrastructure to task the human annotation and to evaluate its quality. For most scholarly IR systems that deal with finding similar documents in large collections, such as scholarly IR systems, this task is often very costly and infeasible to perform mostly due to the size of the documents, the specifics of the domain as well as the domain expertise. Yet these collections often have a need of using a document similarity model either as a retrieval feature such as a document recommendation or as means for performing exploratory data analysis. Deciding which document similarity model to initially use remains an open question which is usually answered by a certain intuition on the system developer side without any quantitative justification. We should note here that once a model is implemented and clickthrough data is collected the similarity model implemented initially could be reevaluated but the question still holds as to which model or model configuration should be first implemented. 

To that end we developed an approach that streamlines the evaluation of document similarity models in large scholarly publication collections. Our approach is solely based on the notion that when retrieving similar documents, the power of the retrieval model should be such that dissimilar documents would be assigned with lower similarity while similar documents should have higher similarity values. Focusing on the precision aspect of the model, more powerful models would have higher precision which implies that similar documents would need to have higher similarity values in order to be ranked higher. As the rank of the similar documents increases we would expect to see their mass on the histogram of similarity values to shift towards minimum values\footnote{As stated earlier for the Cosine distance we would expect to see the opposite unless their values are normalized by their relative range.}. Same notion applies for dissimilar documents - as the precision of the model increases we should expect to see their similarity values to decrease or remain in the same region which in turn implies that their histogram mass would shift towards lower values or assume the same range of values across different models. In our case we threat randomly generated document pairs as dissimilar documents. 

Reflecting on the position of the documents in the shared space, more powerful models should position similar documents close to each other while dissimilar documents should be placed further apart. We compute the ratio between the histograms of the similar and random document pairs across all bins which gives us the proportion of similar documents that we would expect to find at certain similarity values. In the shared space, for a given query document, this ratio gives us the proportion of similar documents that we would expect to find at certain distances. Across different models, as the performance of the model improves the proportions of similar documents found close to the query document should increase thus making the slope of the histogram analysis steeper going down from left to right. We define the absolute value of the slope as an evaluation metric which we call random Histogram Slope Analysis (rHSA). Unlike previous approaches that compare scores of human annotated sets of relevant and non-relevant documents \cite{Krstovski:2015} in our approach the comparison is done over automatically generated sets of consecutively downloaded and randomly generated document pairs. Furthermore in our analysis, as we will show in the next chapter,  we also model the error in the observed similarity values which helps us obtain more accurate slope values.

\section{Extracting Consecutively Downloaded Documents}

In this paper we utilize download logs obtained from the SAO/NASA Astrophysics Data System (ADS) \cite{Kurtz} which is a large scholarly publication retrieval system that covers the fields of Astronomy, Astrophysics and Physics. We mined logs from the past three years and focused our attention to articles from the Astrophysical Journal (ApJ)\footnote{http://en.wikipedia.org/wiki/The\_Astrophysical\_Journal}. We were able to extract a total of ~23M downloads and account for a total of ~1.7M unique downloads. We filtered out consecutively downloaded document pairs that occurred within a time window of more than 10 seconds and less than one hour. These heuristics attempt to eliminate document download pairs that are too short to be generated by humans and too long to belong to the same session. We obtained a total of 3,898,994 consecutively downloaded document pairs which contained 90,395 unique documents. Using the extracted document pairs we created two test sets:
\begin{enumerate}
\item Our first set consists of one million consecutively downloaded document pairs, chosen randomly from the above total set and one million randomly generated document pairs using the 90,395 unique documents. As we will show in the next section we used these two sets of document pairs to introduce rHSA which allows us to predict the performance of different document similarity models. We refer to this set as the 1M set (1M).
\item Using each of the 90,395 unique documents we computed statistics in regards to how many times documents were downloaded immediately after and based on these statistics we created a query set of 100 documents where each document had more than 100 succeeding downloads. In other words the query sets consists of documents where after one of the query documents has been downloaded users have downloaded more than 100 other unique documents. For each of the 100 documents in the query set we considered as pseudo-relevant documents that were consecutively downloaded after the query document more than 10 times. In later sections we refer to this set as the 100 query set (100q). The purpose of this set is to perform traditional IR evaluations over different document similarity models and compare their performance using MAP. 
\end{enumerate}
\section{Document Similarity Models}
We introduce rHSA using two different document similarity models that represent documents in two different spaces - the topic space which is essentially the probability simplex and the vector space which is an instance of the metric space.

We use Vowpal Wabbit \cite{Langford} implementation of LDA which utilizes the online variational Bayes approach \cite{Hoffman} for approximating the posterior per document-topic distributions. This allows us to represent documents in the shared topic space where similarity across documents was computed using JS divergence. Documents were represented using a vocabulary of 36,338 tokens. This vocabulary was generated by analyzing the term frequency values of all tokens occurring in a larger collection of ApJ publications and filtering out tokens whose total frequency count across the collection is less than 10. We also removed the 50 most frequent tokens as well as tokens whose character length is less than four and tokens that have non alphabetic characters such as numeric and special characters. Out of the total set of $\sim$36k tokens we discovered that 16,448 tokens were present in the collection. This constitutes our effective vocabulary. 

For our second approach we used the traditional vector space model \cite{Croft_book} where documents were represented using our effective vocabulary. More specifically, the tf--idf values computed across this vocabulary. We represented each of the non-query documents in the collection (a total of 90,295 such documents) using the tf--idf values across all the tokens in the effective vocabulary. We represented the query documents using four different approaches which we refer to as tf--idf configurations. These configurations vary by the number of top tf--idf values that were used to represent the query document. For a given query document we first compute a ranked list of tf--idf values across the tokens present in the document. We then took the top-n tf--idf tokens (n=50, 100 and 500) to represent the query. Our fourth tf--idf configuration represents the query document using all of its tf--idf tokens. In the vector space model we compute document similarity using the Cosine distance metric. More specifically the normalized Cosine distance: \begin{math}norm.(Cosine)=\frac{Cosine}{max(Cosine)-min(Cosine)}\end{math}.

\section{Computing rHSA}

Computing rHSA requires two steps:
\begin{itemize}
\item \textit{Obtain histograms}: Using a document similarity model we compute similarity across the consecutively downloaded and randomly generated document pairs. We use these similarity values to obtain histograms for the two sets of document pairs. 
\item \textit{Calculate slope of the log division}: In the second step we compute a division between the two histograms in log space. The slope is then computed over the extracted curve from the log division.  
\end{itemize}

\begin{figure}
\centering
\begin{subfigure}{0.72\columnwidth}
\includegraphics[width=\columnwidth]{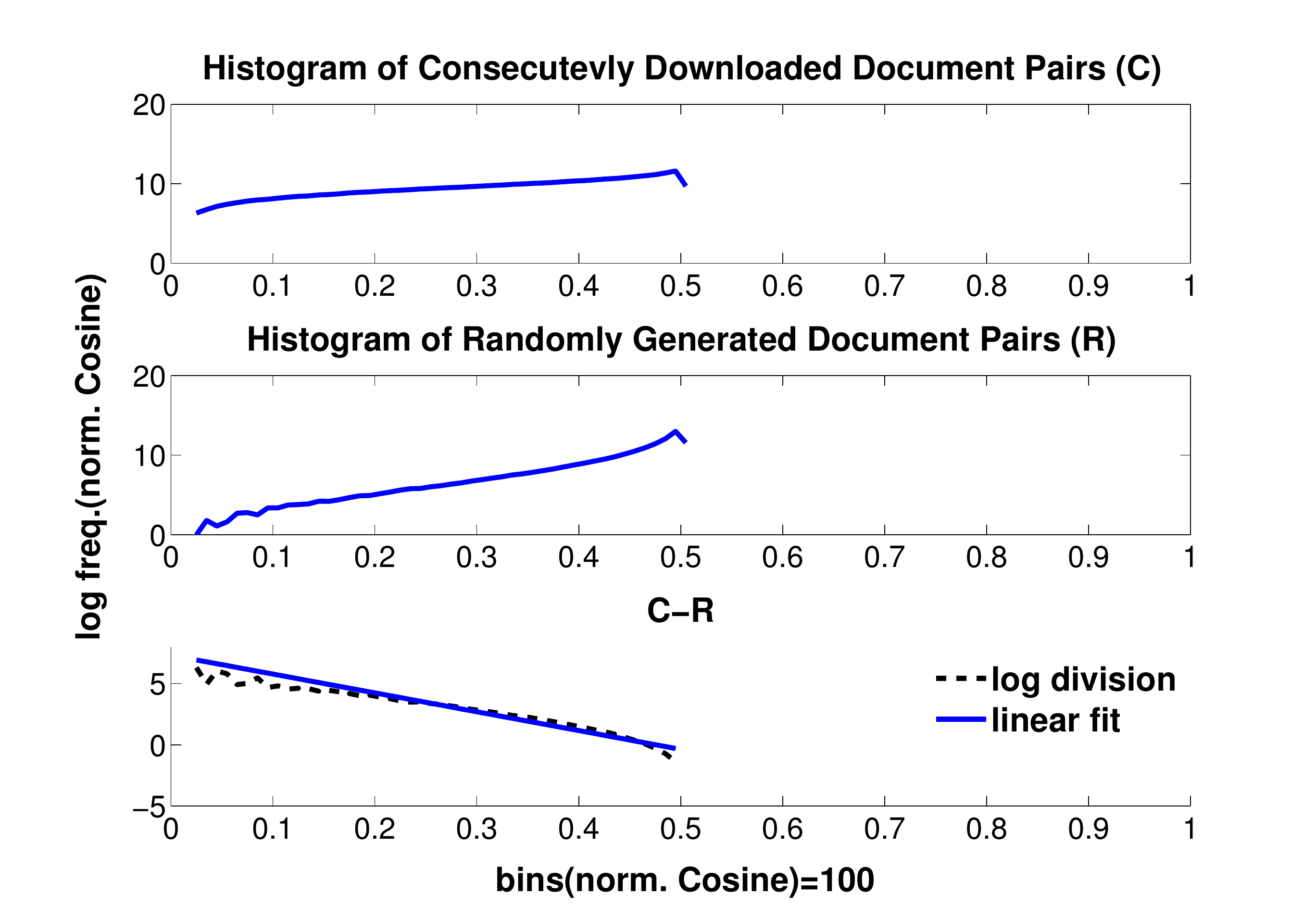}
\caption{}
\label{fig:tfidf_suba1}
\end{subfigure}\hfill
\begin{subfigure}{0.72\columnwidth}
\includegraphics[width=\columnwidth]{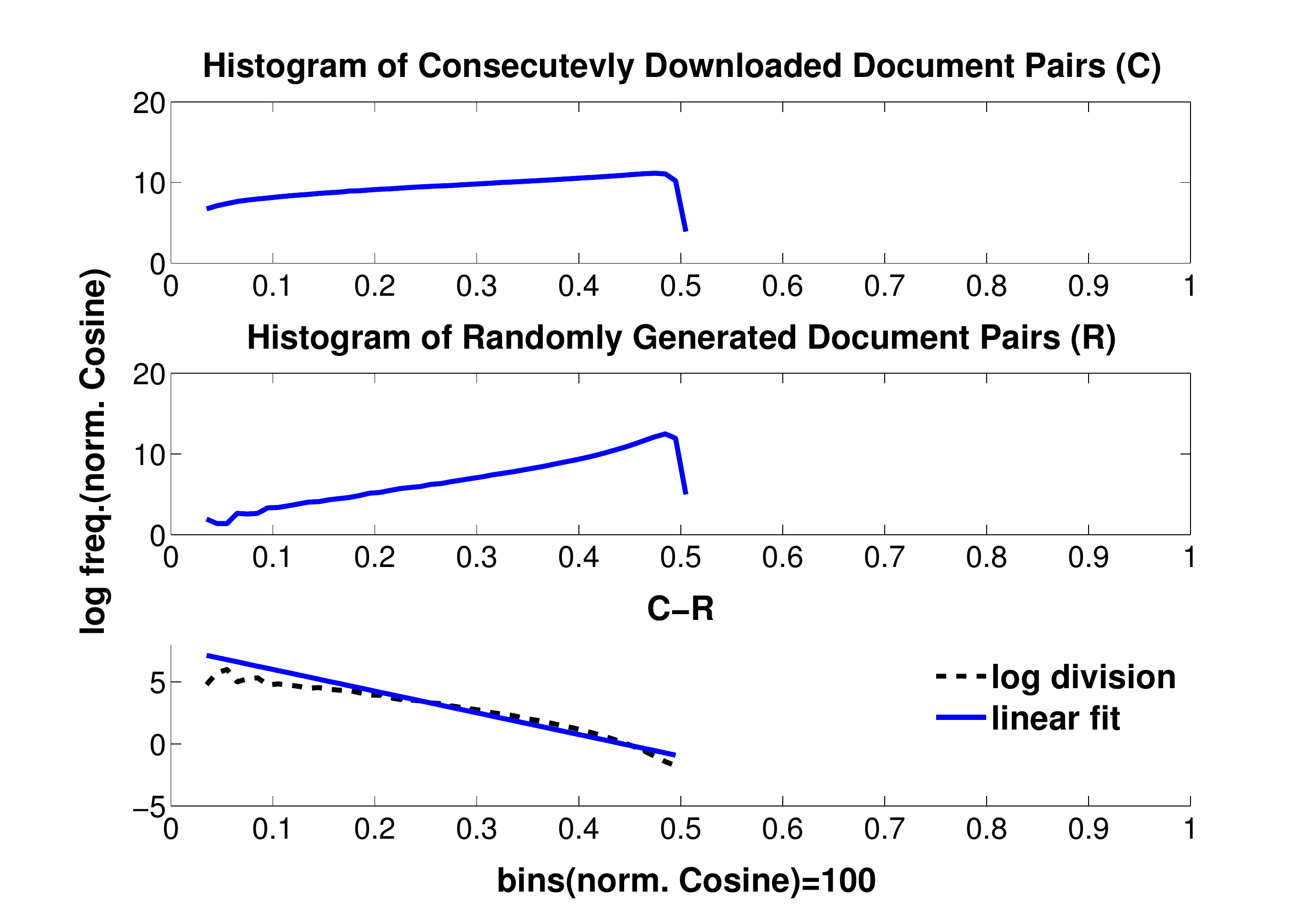}
\caption{}
\label{fig:tfidf_subb1}
\end{subfigure}
\caption{Computing rHSA over 1M consecutively downloaded and randomly generated document pairs. Pairs were obtained using the vector space model where each document was represented using its top 50 tf--idf terms (a) and all tf--idf terms (b). }
\label{fig:logstfidf}
\end{figure}

\begin{figure}
 \centering
 \begin{subfigure}{0.72\columnwidth}
  \includegraphics[width=\columnwidth]{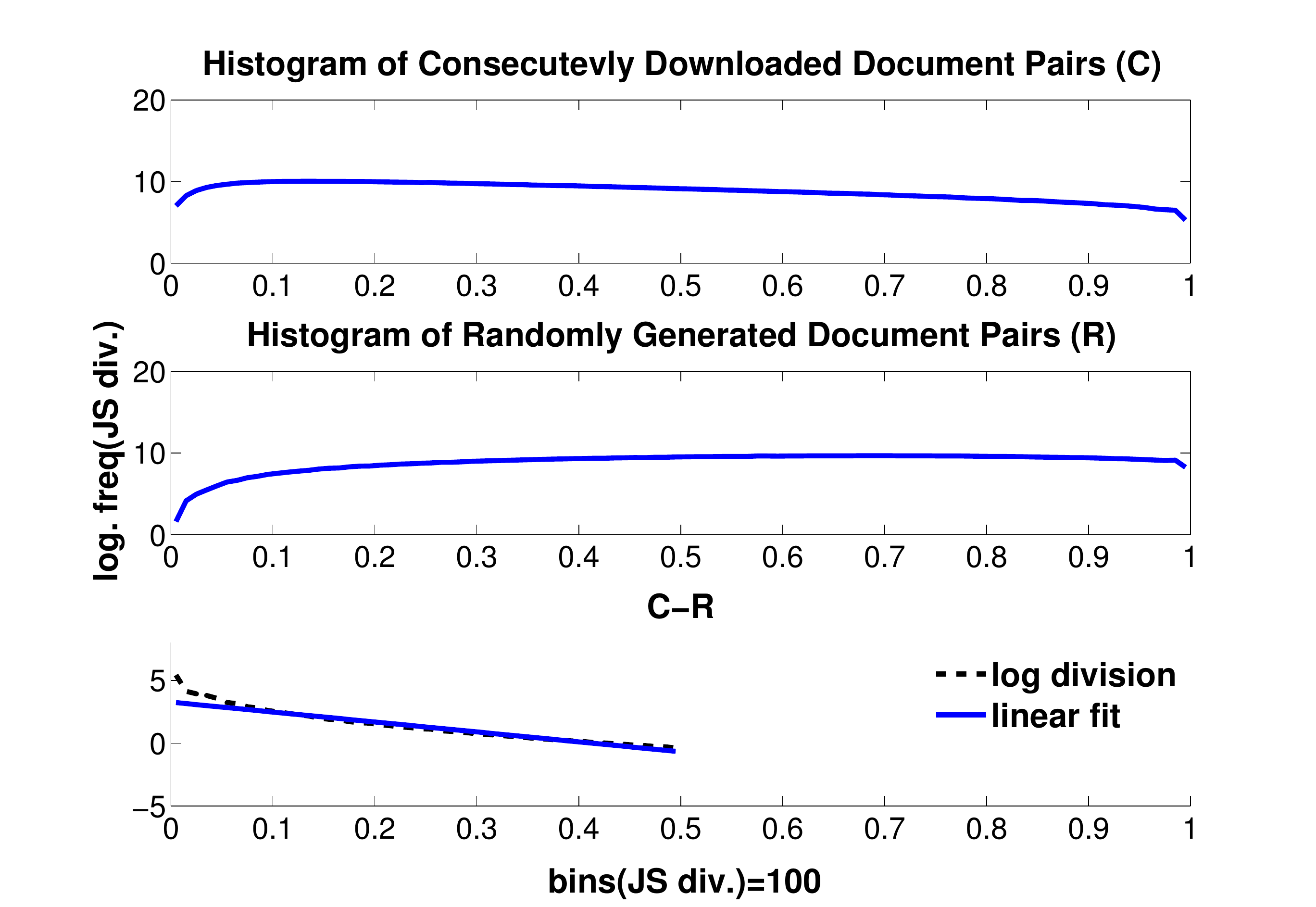}
  \caption{}
  \label{fig:lda_suba2}
 \end{subfigure}\hfill
 \begin{subfigure}{0.72\columnwidth}
  \includegraphics[width=\columnwidth]{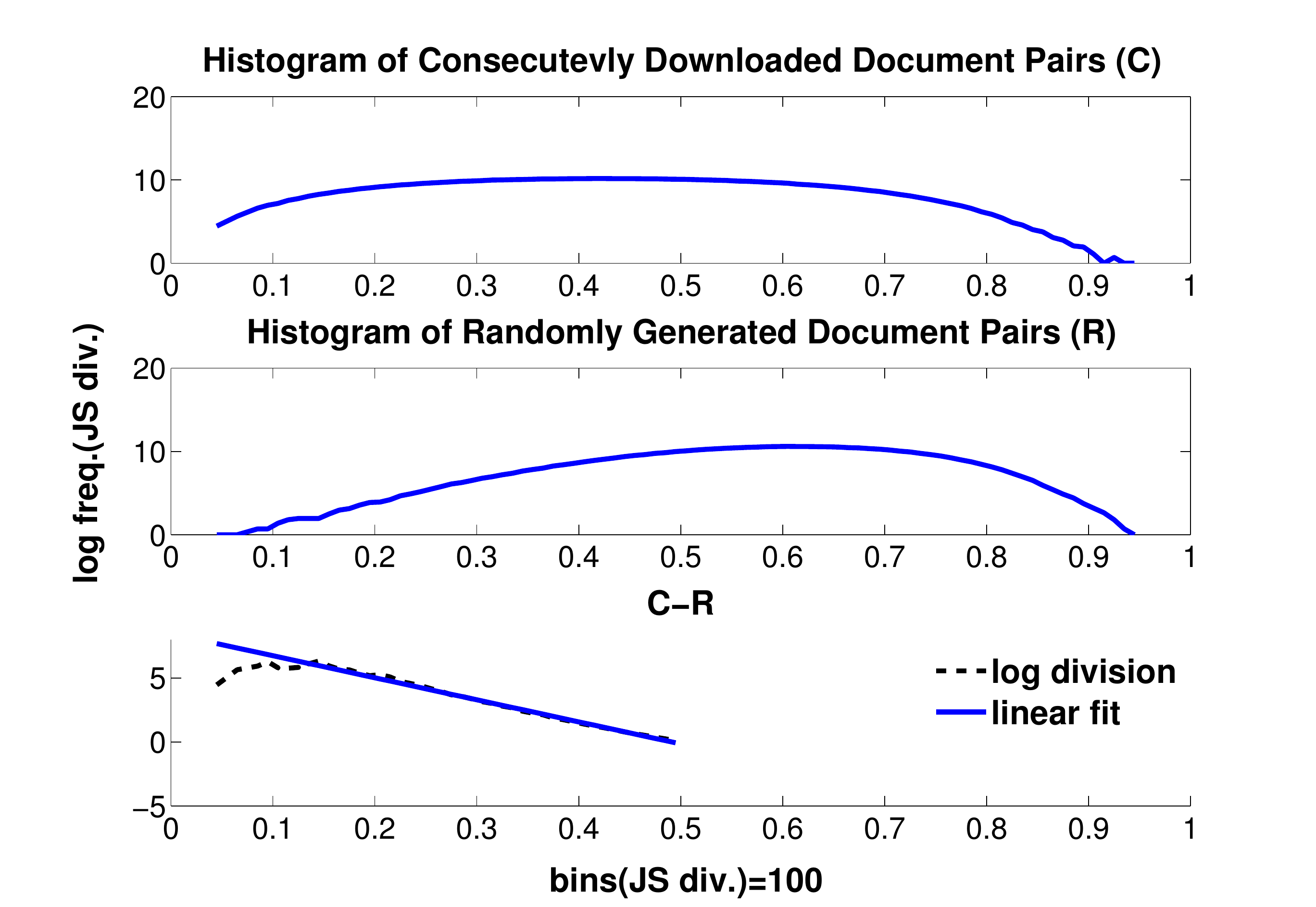}
  \caption{}
  \label{fig:lda_subb2}
  \end{subfigure}
 \caption{Computing rHSA over 1M consecutively downloaded and randomly generated document pairs. Pairs were obtained using LDA with topic configuration of T=50 topics (a) and T=2000 topics (b).}
 \label{fig:logslda}
\end{figure}

Figures ~\ref{fig:logstfidf} and ~\ref{fig:logslda} show examples of computing rHSA for evaluating two configurations of the tf--idf and LDA models. In all figures the first subplot shows the log histogram plot over the normalized Cosine distance and JS divergence values computed across the 1M consecutively downloaded document pairs (C). The second subplot shows the log histogram plot across the similarity values of the 1M randomly generated document pairs (R). The third subplot gives us the result of the log division  ($LogDiv=C-R$) between the two histograms along with the linear fit. Rather than using algebraic operations such as subtraction and division which cannot directly be used to compute the slope due to the fact that their values reflect the difference in the total number of similarity values in the two sets we use log division. This operation smooths out the absolute difference in the total number of values and as such is not directly affected by the size of the two sets of values. 
When computing the histograms we also account for the fact that for each bin the error in the observations of similarity values follows a Poisson distribution. This helps us estimate the error variance. For the Poisson distribution the variance is equivalent to the mean which in our case are the observed counts. For each bin we also model the error in the observed similarity values as a Gaussian distribution whose variance is easy to compute from the actual similarity values that fall within. This allows us to propagate the uncertainties, i.e. errors, in our histogram measurements through the log division operation across both axis. For the error in the observations we have: $\sigma^{2}_{LogDiv_{y}}\:\mathtt{=}\:\frac{\sigma^{2}_{Div_{y}}}{Div^{2}_{y}}$, where $\sigma^{2}_{Div_{y}}\:\mathtt{\simeq}\:Div^{2}_{y}\left(\frac{\sigma^{2}_{C_{y}}}{C^{2}_{y}}+\frac{\sigma^{2}_{R_{y}}}{R^{2}_{y}}\right)$ assuming that for each bin the fluctuations in the observations of the similarity values and their counts are uncorrelated. The slope is then computed using linear regression with errors. 

Across all models, the log division of the two histograms observes an exponential form. For a given query document this functional form, as explained earlier, gives us the distribution of observing similar documents at certain similarity values. More specifically, for the JS divergence and the normalized Cosine distance the distribution could be parametrized with single parameters $\alpha$ and $\beta$: 
\begin{gather*}\label{eq:js}
LogDiv_{JS}\approx\exp^{\alpha*JS}\\
LogDiv_{norm.(Cosine)}\approx\exp^{\beta*norm.(Cosine)}
\end{gather*}

Compared to traditional IR evaluation metrics, rHSA is more efficient to compute. While both approaches do require pseudo-relevant sets of similar documents in case of rHSA there is no need of traversing query results and determining the position of the relevant documents. 

\section{Evaluating Document Similarity Models}

In this section we show the performance of rHSA as a model evaluation metric across two different models and across different model configurations in order to rank and predict the model performance using the 1M set. We further show how rHSA relates to MAP. For this purpose we ran experiments on the traditional IR task where we measured MAP over the 100 query documents and their pseudo-relevant document sets.

\subsection{Model Performance Comparison using rHSA and MAP}\label{hsa_map}

One of the purposes of developing rHSA was to be able to evaluate and predict the performance of document similarity retrieval models without the need of performing traditional IR evaluation task. To demonstrate this ability we computed rHSA over the sets of 1M document pairs and compared its values with MAP computed over the 100 query set. Tables~\ref{tab:map_hsa_tfidf} and ~\ref{tab:map_hsa_lda} show the comparison across different tf--idf and LDA model configurations. In both tables MAP was computed using an evaluation set of 100 queries while rHSA was computed using the query pseudo-relevant and irrelevant documents (100q) and 1M consecutively downloaded and randomly generated document pairs. For LDA with 10k topics the inference procedure generates close to uniform per document topic distributions across both collections which in turn makes all document pairs very similar. Log division over such histograms is zero. 

\begin{table}
  \centering
  \begin{tabular}{|l |r| r |r|}
	 \hline
    Model & MAP(100q) & rHSA(100q) & rHSA(1M)\\
		\hline
		tf--idf top=50 & 0.227 & 16.70 & 15.37\\ 
		tf--idf top=100 & 0.240 & 16.98 & 16.29\\ 
		tf--idf top=500 & 0.248 & 17.30 & 17.47\\ 
		tf--idf all & 0.248 & 17.43 & 17.48\\ 
		\hline
  \end{tabular}
  \caption{Comparison between MAP and rHSA computed across various tf--idf configurations.}
  \label{tab:map_hsa_tfidf}
\end{table}

\begin{table}
  \centering
  \begin{tabular}{|l |r |r |r|}
	 \hline
    Model & MAP(100q) & rHSA(100q) & rHSA(1M)\\
		\hline
		LDA T=50 & 0.077 & 7.54 & 7.89 \\
		LDA T=100 & 0.100 & 9.01 & 8.78\\
		LDA T=500 & 0.221 & 16.33 & 12.12\\
		LDA T=1k & 0.272 & 18.03 & 14.46\\
		LDA T=2k & 0.306 & 21.46 & 17.21\\
		LDA T=3k & 0.297 & 22.45 & 18.01\\
		LDA T=4k & 0.293 & 18.78 & 18.17\\
		LDA T=5k & 0.302 & 20.66 & 18.18\\
		LDA T=10k & 0.005 & 0.0 & 0.0\\
		\hline
  \end{tabular}
  \caption{Comparison between MAP and rHSA computed across various LDA configurations.}
  \label{tab:map_hsa_lda}
\end{table}

For our LDA model configurations, as we increase the number of topics we see a gain in MAP and we obtain the best performance with T=2k. Further increasing the number of topics we observe a drop in the model performance. In case of the tf--idf model, increasing the number of top tf--idf words we also obtain an improvement in MAP up to top=500. Further increasing the number of top tf--idf tokens doesn't provide us with any additional improvements over MAP. Across the two models rHSA evaluates LDA with T=5k as the best overall model and best LDA configuration. 
In case of the vector space rHSA picks the tf--idf configuration with all tokens as its best configuration. 

We also compared rHSA over the collection of 100 query documents where the set of consecutively downloaded document pairs consisted of the query and pseudo-relevant documents and the randomly generated documents pairs consisted of the query and non-relevant documents. Shown in the third column of Tables ~\ref{tab:map_hsa_tfidf} and ~\ref{tab:map_hsa_lda} is the comparison across different tf--idf and LDA model configurations over the set of 100 queries. When computed over the 100 query set rHSA gives the best result for the LDA model with T=3k. In this case the performance of all tf--idf models is ranked between the performance of LDA with T=500 and T=1k which correlates well with the rank of the performance of these models using MAP. 

\subsection{Predicting Model Performance using rHSA}

We evaluated the predictive power of rHSA by performing linear and rank correlation between the ranked list of the models' performance sorted by MAP and the ranked list obtained using rHSA. Linear correlation was computed using Pearson correlation coefficient (R) and for rank correlation we used Spearman's coefficient ($\rho$). Table ~\ref{tab:rsqr} shows the correlation coefficients across the two families of models and across the two collections that we use to compute rHSA. 

\begin{table}
  \centering
  \begin{tabular}{|l |r |r| r|}
	 \hline
    Doc. Sim. Model Type & Collection & $R$ & $\rho$\\
		\hline
		\multirow{2}{*}{tf--idf} & 100q & 0.97  & 1.00 \\ 
		\multirow{2}{*}          & 1M   & 0.98  & 1.00 \\ 
		\hline
		\multirow{2}{*}{LDA}     & 100q & 0.99 &  0.95 \\ 
		\multirow{2}{*}          & 1M   & 0.97 &  0.88 \\ 
		\hline
  \end{tabular}
  \caption{Linear ($R$) and rank ($\rho$) correlation coefficients values across different linear fits between rHSA and MAP computed over the two model families. Shown in the second column are the collections used to compute rHSA. }
  \label{tab:rsqr}
\end{table}

Across both family of models and rHSA compute collections, rHSA yields a very high linear and rank correlation with MAP thus giving rHSA a very high predictive power of the model performance. Across the two query sets we achieve almost the same linear and rank correlation on the tf--idf family of models. In case of the 1M query set higher linear and rank correlation is achieved over the 100q query set. In practice the high correlation coefficients across both models and query sets allow us to compute rHSA on a set of consecutively downloaded and randomly generated document pairs and predict the performance of the document similarity model prior to using that model on a particular document similarity task over the same collection.

\section{Conclusion}

Evaluating document similarity models traditionally requires a test collection consisting of a set of query documents and their relevant document sets. We presented a novel approach for automatically generating document similarity evaluation sets by mining download logs and utilizing information on consecutively downloaded documents extracted from scholarly IR system. We showed that model evaluation could be achieved using such a document collection. Furthermore we presented rHSA - a new approach that streamlines the performance evaluation of document similarity models in large collections of scholarly publications. We showed that rHSA achieves high correlation with MAP. Finally we showed that rHSA offers the ability to predict the performance of the document similarity model without the need of a typical IR evaluation set.

\section{Acknowledgment}
This work was supported in part by the Center for Intelligent Information Retrieval. Any opinions, findings and conclusions or recommendations expressed in this material are those of the authors and do not necessarily reflect those of the sponsor.

\bibliographystyle{modified}
\fontsize{9.5pt}{10.5pt}\selectfont
\bibliography{Krstovski_Smith_Kurtz} 
\end{document}